\pdfoutput=1

\documentclass[runningheads,a4paper]{llncs}[2015/06/24]
\usepackage{makeidx}

\usepackage[utf8]{inputenc}
\usepackage{cmap}
\usepackage[T1]{fontenc}

\usepackage[strings]{underscore}
\usepackage{rotating}
\usepackage{chngcntr}
\usepackage{listings}
\usepackage[table]{xcolor}
\usepackage[normalem]{ulem}

\usepackage{graphicx}

\usepackage{amsmath}
\usepackage{amssymb}
\usepackage{url}
\usepackage{color}

\usepackage[%
rm={oldstyle=false,proportional=true},%
sf={oldstyle=false,proportional=true},%
tt={oldstyle=false,proportional=true,variable=true},%
qt=false%
]{cfr-lm}

\usepackage{cite}

\usepackage{paralist}
\usepackage{csquotes}
\usepackage{microtype}

\usepackage{url}
\makeatletter
\g@addto@macro{\UrlBreaks}{\UrlOrds}
\makeatother

\usepackage{pdfcomment}
\hypersetup{hidelinks,
   colorlinks=true,
   allcolors=black,
   pdfstartview=Fit,
   breaklinks=true}
\usepackage[all]{hypcap}

\usepackage[capitalise,nameinlink]{cleveref}
\crefname{section}{Sect.}{Sect.}
\Crefname{section}{Section}{Sections}

\usepackage{xspace}
\newcommand{\eg}{e.\,g.,~}
\newcommand{\ie}{i.\,e.,~}

\definecolor{bluekeywords}{rgb}{0.13,0.13,1}
\definecolor{greencomments}{rgb}{0,0.5,0}
\definecolor{redstrings}{rgb}{0.9,0,0}

\usepackage{enumitem}
\setlist[itemize]{noitemsep, topsep=0pt}
\setlist[enumerate]{noitemsep, topsep=0pt}

\let\llncssubparagraph\subparagraph
\let\subparagraph\paragraph
\usepackage{titlesec}
\let\subparagraph\llncssubparagraph

\titlespacing{\section}{0pt}{10pt plus 0pt minus 2pt}{8pt plus 0pt minus 2pt}
\titlespacing{\subsection}{0pt}{8pt plus 0pt minus 1pt}{6pt plus 0pt minus 1pt}

\begin{document}

\counterwithout{lstlisting}{chapter}

\setlength{\abovedisplayskip}{1pt plus 1pt}
\setlength{\belowdisplayskip}{1pt plus 1pt}

\title{Deductive Verification of Unmodified Linux Kernel Library Functions}

\author{Denis Efremov\inst{1,2} \and Mikhail Mandrykin\inst{2} \and Alexey Khoroshilov\inst{1,2,3,4}}
\institute{National Research University Higher School of Economics, Moscow, Russia\\
\and
Ivannikov Institute for System Programming of the RAS, Moscow, Russia\\
\and
Moscow Institute of Physics and Technology, Moscow, Russia\\
\and
Lomonosov Moscow State University, Moscow, Russia\\
\email{defremov@hse.ru, \{mandrykin,khoroshilov\}@ispras.ru}
}

\maketitle

\begin{abstract}
This paper presents results from the development and evaluation of a deductive verification benchmark consisting of 26 unmodified Linux kernel library functions implementing conventional memory and string operations. The formal contract of the functions was extracted from their source code and was represented in the form of preconditions and postconditions. The correctness of 23 functions was completely proved using AstraVer toolset, although success for 11 functions was achieved using 2 new specification language constructs. Another 2 functions were proved after a minor modification of their source code, while the final one cannot be completely proved using the existing memory model. The benchmark can be used for the testing and evaluation of deductive verification tools and as a starting point for verifying other parts of the Linux kernel.
\end{abstract}


\setcounter{footnote}{0}

\section{Introduction}

Deductive verification is one of the most rigorous techniques to ensure software satisfies its requirements. In spite of significant advances in tool support, it still requires deep user involvement in the verification process to provide manual guidance (\eg to specify the contract of each function and to identify loop invariants). As a result deductive verification is used mainly to analyze the most critical pieces of software.

Under such conditions, it is more cost-effective to rewrite code to make it easier to verify than to implement support for all the complex corner cases in the semantics of the target programming platform, most notably in low-level platforms based on C which lack well-defined semantics for many cases widely used in practice.

Nevertheless, there are situations where changing the code under verification is undesirable or even impossible. For example, components to be integrated into a predefined framework have to follow the coding style, interfaces and data structures of that framework. We have met such limitations with Linux kernel modules where a lot of implementation details are imposed by Linux kernel core infrastructure.

One of the well-established approaches to specifying the behavioral contract of functions written in C is ANSI/ISO C Specification Language (ACSL)~\cite{acsllang}. \textsc{Frama-C}~\cite{framac} provides a framework for an analysis of C programs with optional ACSL specification annotations. \textsc{Frama-C} integrates specifications and code into a single intermediate representation and allows plugins to work with it. There are two plugins for deductive verification built on top of \textsc{Frama-C}: \textsc{WP}~\cite{framac} and \textsc{Jessie}~\cite{moyjessiephd}.

Because existing plugins were not able to correctly handle many constructs widely used in the Linux kernel (\eg \texttt{container_of}, pointer type reinterpretation between integer types of different size), we started developing a new deductive verification \textsc{AstraVer} plugin based on \textsc{Jessie}. We implemented and proposed many improvements for the toolchain, including a new memory model~\cite{jessiemodel}, but there is no representative benchmark to evaluate the progress. The primary purpose of this work is to fill this gap.

Following previous efforts \cite{klibcverification,openbsdverification}, we have chosen for the first step Linux kernel library functions implementing conventional memory and string operations. The benchmark built from such functions helped us to detect a number of local tool issues and several fundamental problems discussed below.

The main contributions of the paper include:
\begin{itemize}
  \item a benchmark of unmodified Linux kernel library functions extended with annotations formalizing their contracts in the form of preconditions and postconditions~\cite{url:verkerresults};
  \item a new approach to annotate modulo arithmetic operations on values of integral C types;
  \item evaluation of \textsc{AstraVer} deductive verification toolchain on the benchmark.
\end{itemize}

The paper is organized as follows. Section~\ref{sec:related} discusses similar efforts aimed at the specification and deductive verification of C library functions. Section~\ref{sec:acsl} provides background on ACSL basic concepts. Section~\ref{sec:toolsupd} presents improvements in the toolchain made during the development of the benchmark. Sections~\ref{sec:specs} describes specification techniques designed and applied for specification of library functions. Section~\ref{sec:issues} defines a set of open problems. Section~\ref{sec:evaluation} presents the evaluation of the solvers. Section~\ref{sec:conclusion} summarizes the results of the work.

\section{Related Work}\label{sec:related}

Since the deductive verification tools, \textsc{WP} and \textsc{Jessie} are mature enough there are many examples where these tools were applied for verification of real-life software. In~\cite{openbsdverification} 12 string functions from OpenBSD were examined, using \textsc{Jessie} as a deductive verification plug-in. The correctness of 7 functions was fully proved (all verification conditions, or VCs, were successfully discharged). For the other 5 functions, some VCs were left unproved. The author did three iterations on the development of a specification contract for each function. First, one was developed based on the standard and the author's experience. The second one was developed based on informal documentation (man pages) exclusively. The final one was written based on the implementation (source code) and the man pages. The final revision in most cases has significant differences from previous versions. Thus it shows that it is difficult to develop a formal specification in ACSL language for already developed source code without taking the implementation into account. However, such an iterative approach allowed the author to find inaccuracies in the documentation for several functions, and a lack of documentation completeness in many cases.

To prove some functions, the author changed the source code. Changes were performed in two specific situations. In the first case \texttt{char *} type in \texttt{strcmp} and \texttt{strncmp} functions were cast to \texttt{unsigned char *}. In the second case, the unsigned loop iterator in \texttt{strlcat} underflowed at the last iteration step due to the postfix decrement. The loop termination, in this case, occurs when the variable equals zero, but after comparison, the value of the variable is still decreasing by one. This results in the unsigned integer underflow (which is not an undefined behavior). However, the unsigned underflow does not lead to an error in the code: after the loop, the variable is not used anywhere. But in this situation, it is not possible to prove the VC demanding the absence of an integer overflow (more generally, over- or underflow). The VC is necessary due to the use of the \texttt{defensive} integer model (see section \Cref{section:combint}) when the bounded integers are modeled using mathematical unbounded integers assuming the absence of integer overflows. To prove the VCs, \textsc{Alt-Ergo} (0.7.3), \textsc{Simplify} (1.5.4) and \textsc{Z3} (2.0) solvers were used.

In~\cite{klibcverification} authors used \textsc{Frama-C} with the \textsc{WP} deductive verification plugin to verify the functions of the \textsc{klibc} library. The authors were able to fully prove 14 string functions. For 12 functions some VCs were not discharged. Four more functions failed to analyze due to errors in the verification tools. In addition to the string functions (from \texttt{string.h}), functions from the \texttt{stdio.h} were also analyzed. As noted by the authors, almost all functions from this header file use system calls, which in most cases results in a weak specification. To overcome the limitations of the verification tools and to simplify the generated VCs the authors made changes to the source code.

The authors analyzed in advance the problems with type casts modeling (for example, \texttt{unsigned char *} to \texttt{char *}) and modified the code to exclude such operations. The authors also faced the code pattern with the postfix decrement in a while loop. To prove the VCs, \textsc{Alt-Ergo} (0.95.1), \textsc{CVC3} (2.4.1) and \textsc{Z3} (4.3.1) solvers were used.

The most comprehensive document on ACSL specifications development is ACSL by Example~\cite{acslbyexample}. It contains ACSL specifications for functions from the C++ standard library (Standard Template Library). Initial implementation converted from C++ function templates to C functions that work on arrays of type \texttt{int}. The authors regularly update the document with new specifications and functions, bug-fixes, etc. This project started in 2009. The document contains a number of fully verified functions. They were proved with \textsc{Alt-Ergo}, \textsc{CVC3}, \textsc{CVC4}, \textsc{Z3}, and \textsc{EProver} solvers. Authors use the \textsc{WP} deductive verification plugin.

GrammaTech report~\cite{grammatechverification} describes typical problems the authors encountered when developing specifications for the \textsc{GTlibc} library. \textsc{Frama-C} with \textsc{WP} was used. Among other points, the authors report memory model problems with pointers type casts and pointers comparison.

\section{ACSL}\label{sec:acsl}

ACSL is designed to be suitable for specifying safety properties of C programs, including contract specifications (pre- and postconditions) and assertions with arbitrary predicates on one or several memory states. The language also supports the specification of function frame conditions, axiomatic theories and additional annotations required by particular verification tools to check the specified properties (examples of additional specifications are loop invariants and pragmas). ACSL includes specification constructs for expressing C-specific attributes related to explicit low-level memory management such as start addresses and lengths of allocated memory blocks, pointers with support for arbitrary pointer type conversions and accessibility predicates for read-only and read-write access. 

\begin{lstlisting}[caption={from Linux 4.12, \texttt{lib/string.c}},captionpos=b,label={lst:strnchr}]
/*@ requires valid_strn(s, count);
    assigns \nothing;
    behavior exists:
     assumes ∃ char *p; s≤p<s+strnlen(s,count) ∧ *p ≡ (char %)c;
     ensures s ≤ \result ≤ s+strnlen(s, count);
     ensures *\result ≡ (char %) c;
     ensures ∀ char *p; s ≤ p < \result ⇒ *p ≢ (char %)c;
    behavior not_exists:
     assumes ∀ char *p; s ≤ p < s+strnlen(s,count) ⇒ *p ≢ (char %)c;
     ensures \result ≡ \null;
    complete behaviors;
    disjoint behaviors;*/
char *strnchr(const char *s, size_t count, int c) {
 //@ ghost char *os = s;
 //@ ghost size_t ocount = count;
 /*@ loop invariant 0 ≤ count ≤ ocount;
     loop invariant os ≤ s ≤ os+strnlen(os, ocount);
     loop invariant s-os ≡ ocount-count;
     loop invariant valid_strn(s, count);
     loop invariant strnlen(os,ocount) ≡ s-os+strnlen(s, count);
     loop invariant ∀ char *p; os ≤ p < s ⇒ *p ≢ (char %) c;
     loop variant count;
  */
 for (; count-(*@\,@*)-/*@%*/ && *s != '\0'; ++s)
  if (*s == (char)/*@%*/c)
   return (char *)s;
 return NULL;
}
\end{lstlisting}

\sloppypar
Let's consider an example of a simple C function with an appropriate ACSL specification. \Cref{lst:strnchr} presents one of the implementations for function \texttt{strnchr} from the Linux kernel. The function searches for the first occurrence of character \texttt{c} in a string \texttt{s} of length bounded by the parameter \texttt{cnt}. The pre-condition in line \texttt{1} requires the string \texttt{s} to address a valid memory area of length \mbox{\texttt{min(strlen(s), cnt) + 1}}. \texttt{strnchr} is a pure function, the absence of effects on memory state is specified in line \texttt{2}. The further specification is split into two cases: The first one when the string includes the searched character and the second one when it does not. ACSL includes a special construct for such composite specifications, which is called \textit{behaviors}. In ACSL behaviors are not treated as syntactic sugar (unlike, \eg JML), but fully integrated into the language such that nearly all specification constructs both in contracts and in function bodies are attributed to one or several behaviors and thus different behaviors of a function are intended to be checked separately. To verify the function \texttt{strnchr} against its contract specification with a deductive verification tool, the loop invariant and a ranking function (loop variant) are specified in lines \texttt{16}--\texttt{22}.

The implementation of \texttt{strnchr} contains an intentional type cast \texttt{(char)c} in line \texttt{25} and a postfix decrement of a loop iterator \texttt{count} in line \texttt{24}. In both of those cases, the corresponding operation (type cast or decrement) discards some parts of the bitwise representation of the argument (higher bits of the \texttt{int} value and the sign bit correspondingly), which corresponds to the intention of the programmer. To distinguish those intentionally overflowing operations, whose semantics is described in terms of bitwise interpretation of bounded integers, we introduced a special annotation construct \texttt{/*@\%*/}. 


\section{Region separation in \textsc{Jessie}}\label{sec:region}
Since there are two deductive verification plugins for the \textsc{Frama-C} platform, we had to make a choice between \textsc{Jessie} and \textsc{WP}. While there may be many arguments for choosing a more up-to-date and actively maintained \textsc{WP} plugin, which, among others, has capabilities for bitwise modelling of in-memory data representation and support for interactive proofs, here we emphasize that our initial justification for choosing \textsc{Jessie} over \textsc{WP} was due to its more flexible architecture that enabled easier experimentation with custom ACSL extensions (including the composite integer model described in section~\Cref{section:combint}) and also its support for region-based modelling of the heap.

In particular, the heap in \textsc{Jessie} is separated into disjoint regions according to the results of a preliminary conservative static analysis presented in~\cite{regionmarche}. While the separation analysis is coarse (so that its soundness is easy to establish), it is still useful in many cases arising during verification of imperative code. For example, consider the following loop invariant:

\begin{lstlisting}[caption={from Linux 4.12, \texttt{lib/string.c}},captionpos=b,label={lst:strcat}]
/*@ loop invariant \at(src,Pre) ≤ src ≤
                     \at(src,Pre)+strlen(\at(src,Pre));
    loop invariant \at(dest,Pre) ≤ dest ≤
                     \at(dest,Pre)+strlen(\at(src, Pre));
    //...
*/
while ((*dest++ = *src++) != '\0')
  ;
\end{lstlisting}
  Here, in general, proof of the fact that \texttt{strlen\{LoopCurrent\}(\textbackslash{at}(src, Pre)) == strlen\{Here\}(\textbackslash{at}(src, Pre))} holds at the end of every iteration requires inductive reasoning since the definition of \texttt{strlen} is recursive and the side effect of the assignment \texttt{*dest++ = *src++} can generally interfere with the memory footprint of the function \texttt{strlen}. But the  static separation analysis implemented in \textsc{Jessie} assigns disjoint memory regions to the pointers \texttt{dest} and \texttt{src} and so both applications of \texttt{strlen} to \texttt{src} before and after the loop iteration are encoded using a heap variable separate from that of \texttt{dst} and therefore literally coincide. So in \textsc{Jessie} the non-interference trivially holds and does not require any additional proof effort. In general, the separation analysis is imprecise and may require explicit weakening, \eg if the surrounding function can be called in context with more aliasing (\eg when \texttt{src} may intersect with \texttt{dest}), but it can still considerably simplify the verification by eliminating the need in inductive framing lemmas.

\section{Limitations of the current implementation}\label{sec:toolsupd}
  \subsection{\textsc{Jessie} byte-level block memory model}
  There are a number of ways to logically represent pointers and memory blocks in the generated VCs. \textsc{Jessie} implements the \emph{byte-level block memory model}~\cite{moyjessiephd}, where pointers are logically represented as pairs of the form $(l,o)$ and memory blocks are represented as triples of the form $(l,a,s)$. Here $l$ is a label uniquely identifying a memory block, $o$ is the offset of the pointer from the starting address $a$ of the block $l$, and $s$ is the size of the block. The introduction of unique block labels allows us to ensure that no memory access occurs beyond the bounds of the pointed memory block even if the corresponding memory area is also allocated. Although such access cannot break segmentation checks, it is forbidden by C standard~\cite{c11} (subsection~{6.5.6}, paragraph~8 classifies out-of-bounds pointers, except for pointers to the one past the last element of an array, as undefined). As explained in~\cite{moyjessiephd} describing the design choices behind the \textsc{Jessie} tool, byte-level block memory model in principle allows us to express common but non-standard C code fragments, such an implementation of the function \texttt{memmove}, while retaining the ability to detect use-after-free memory safety errors and potential pointer overflows.

  The actual implementation of the memory model in the tool, however, diverges from its simple theoretical description in several ways and imposes a number of additional restrictions on the supported subset of C.

  First, pointers are implemented in the corresponding \textsc{Jessie} theory (in WhyML) as values of an abstract type \texttt{pointer} with four corresponding abstract operations:
  \[
    \begin{array}{l}
      \texttt{sub\_pointer}~:~\texttt{pointer}\times\texttt{pointer}\to\texttt{int}, \\
      \texttt{shift}~:~\texttt{pointer}\times\texttt{int}\to\texttt{pointer}, \\
      \texttt{same\_block}~:~\texttt{pointer}\times\texttt{pointer}\to\texttt{bool},~\text{and} \\
      \texttt{address}~:~\texttt{pointer}\to\texttt{int}\text{.}
    \end{array}
  \]
  Block sizes are represented implicitly by so-called \emph{allocation tables}, mutable values of an abstract type with two axiomatically defined functions
  \[
    \begin{array}{l}
      \texttt{offset\_min}~:~\texttt{alloc\_table}\times\texttt{pointer}\to\texttt{int}~\text{and} \\
      \texttt{offset\_max}~:~\texttt{alloc\_table}\times\texttt{pointer}\to\texttt{int}\text{.}
    \end{array}
  \]
  The functions represent the minimal and maximal allowed offset of a pointer in its corresponding allocated memory block \ie for a pointer $p=(l, o)$ and its corresponding memory block $(l,a,s)$ which has size $s$ in the state represented by allocation table $t$, $\texttt{offset\_min}(t,p)=-o$, $\texttt{offset\_max}(t,p)=s-o-1$. Thus a pointer $p$ can be safely dereferenced iff ${0 \le o \le s - 1}$ \ie ${\texttt{offset\_min}(t, p) \le 0} \land {\texttt{offset\_max}(t, p) \ge 0}$. We denote this condition as $\texttt{valid}(t,p)$. There is no direct representation for block labels ($l$) or starting addresses ($a$) of the memory blocks. The VCs generated for dynamic memory allocations and deallocations (function calls to \texttt{kmalloc} and \texttt{kfree} are treated specially in \textsc{Jessie}\footnote{The special treatment is necessary because \textsc{Jessie} does not support arbitrary pointer type casts, in particular, reinterpretation casts such as \texttt{char *}$\to$\texttt{int*}, so the return type of memory allocating functions should be specialized at each call site, which can not be directly expressed in ACSL.}) involve only allocation tables and functions \texttt{sub\_pointer}, \texttt{shift} and \texttt{same_block}. This makes the corresponding axiomatization inherently incomplete. In particular, the function $\texttt{address}$ is not only left entirely uninterpreted in the current implementation, but cannot be even theoretically given a complete axiomatization. Consider the following property of this function: \emph{two valid pointers from different blocks cannot have the same address}. It cannot be expressed as a logical proposition using the current \textsc{Jessie} theory since this would involve bounded existential quantification over all possible \emph{reachable} states of the corresponding allocation table:
  \[
    \begin{array}{l}
      \forall p_1, p_2.~\big(\exists t.~\textit{Reachable}(t) \land \texttt{valid}(t, p_1) \land \texttt{valid}(t, p_2)\big) \land \neg \texttt{same\_block}(p_1, p_2) \\
      ~~~~\quad {} \implies \texttt{address}(p_1) \neq \texttt{address}(p_2)\text{.}
    \end{array}
  \]
  Since the problem of inferring an explicit representation of the predicate $\textit{Reachable}(t)$ is undecidable, the tool should implement an implicit encoding of the pointer address properties at every allocation point:
  \[
    \begin{array}{l}
      \forall p. ~\texttt{valid}(t, p) \\
      ~{} \implies \texttt{address}(p) < \texttt{address}(p^{*}) \lor \texttt{address}(p) \ge \texttt{address}(p^{*}) + \texttt{\textit{sizeof}}(\texttt{*}p^{*}) \times s \text{,}
    \end{array}
  \]
  where $p^{*}$ points to the start of a freshly allocated memory block of size $s \times \texttt{\textit{sizeof}}(\texttt{*}p^{*})$ and $t$ is the state of the allocation table just before the allocation. The unavailability of a precise formalization for the function $\texttt{address}$ prevents the generation of the appropriate VCs for potential pointer overflows and a more flexible formalization of pointer comparison and difference operations (allowing the verifation of functions such as \texttt{memmove}).

Moreover, the pointer offset and difference, as formalized by the functions \texttt{shift} and \texttt{sub\_pointer}, are measured in units equal to the sizes of the addressed values, according to the pointer indexing semantics of C, rather than in bytes or words. In particular, an expression \texttt{p + 1}, where \texttt{p} has type \texttt{\textbf{int} *}, is translated roughly as $p + 1$ rather than as $p + s$, where $s$ is the size of the integer type (usually equal to 4). Such translation immediately prevents many common combinations of pointer casts and arithmetic, including the uses of the \texttt{container\_of} macro. To see this, it is enough to consider two pointers: \texttt{p + 1} and \texttt{((\textbf{char} *)p) + 1}, where $p$ has type \texttt{\textbf{int} *} and points to the beginning of an allocated memory block. In the byte-level block memory model with size-proportional offsets, these pointers would have the same representation $(l, 1)$, while their actual addresses cannot be equal (they should differ at least by 1, usually by 3). This contradicts the functional consistency of the function \texttt{address}. To circumvent this contradiction (and for other reasons, see~\cite{regionmarche,unionandcast}) current \textsc{Jessie} implementation makes use of two separate techniques. First, it introduces \emph{tag tables} tracking the precise dynamic types of the objects in the allocated memory. These tag tables allow us to introduce the necessary checks for pointer shift operations in the generated safety VCs (more on this in~\cite{unionandcast,jessielow}). Second, it implements a number of \emph{normalizing} code transformations that rewrite nested structures and addressed fields of simple types into pointers to separately allocated structures or values of the corresponding type (the transformations are described in~\cite{regionmarche}). This allows us to express the addresses of nested objects in the \textsc{Jessie} memory model. However, a combination of these two approaches results in a number of significant restrictions. In particular, unions containing nested structures as their members cannot be soundly represented by the model. This is because that it is impossible to approximate statically whether a pointer to a structure obtained, say, as a function parameter is actually a pointer to a structure nested in some union and so writing to a field of this structure should be translated into a \emph{strong coercion}~\cite{unionandcast} of the corresponding outer union possibly invalidating other representations of the underlying memory and updating the tag table.

To address these and some other limitations of the current \textsc{Jessie} memory model, a new model was proposed in~\cite{jessiemodel}. This model, though, suggests simple byte-level modeling of pointers. Since we usually assume an arbitrary memory allocation strategy, this should not lead to missed C standard violations due to the dereferencing of valid pointers in different memory blocks in practice. This is because usually in such cases at least one of the possible arbitrary allocation strategies leads to the dereference of an invalid pointer and thus it is impossible to spuriously prove that such a dereference is safe. However, the memory model suggested in~\cite{jessiemodel} is not yet implemented in the tool. So in this study, we used the current implementation of the \textsc{Jessie} memory model as-is.

The only change we made to the tool concerns the translation of pointer inequalities. Since the current implementation does not provide enough support for arbitrary pointer comparisons, we restricted pointer inequalities to support only pointers in the same memory block by generating the corresponding VCs and changing the translation of the corresponding predicates of the form ${p_1\mathrel{\Diamond}p_2}$ into ${\texttt{sub\_pointer}(p_1, p_2)\mathrel{\Diamond}0 \land \texttt{same\_block}(p_1, p_2)}$ (here $\Diamond \in {\{>}, <, =, \le, \ge, {\neq\}}$). This made many specifications slightly shorter as the pervasive condition $\texttt{same\_block}(p_1, p_2)$ was made implicit.
\vfill
\pagebreak

\subsection{\textsc{Jessie} integer models, composite integer model and modulo arithmetic annotations}\label{section:combint}
\textsc{Jessie} originally implements three logical models for machine integer types of different size and signedness. The simplest model called \texttt{math} (or \emph{unbounded}) unconditionally encodes values of all integer types as mathematical integers. It does not support overflow checks and does not model the wrap-around behavior of machine integers. It in principle allows the modelling of some bitwise operations on unbounded integers with an appropriate axiomatization, but in practice such modeling is usually very inefficient. Another, most commonly used integer model is called \texttt{defensive} (or \emph{bounded}) and differs from the \texttt{math} model in two ways:
\begin{itemize}
  \item for integer operations in code it generates appropriate VCs preventing arithmetic overflows;
  \item bounded integral types in \emph{logic} (\ie in specifications) are modeled by abstract types with special injection/projection functions (\eg~\mbox{\texttt{int32\_of\_integer}}~/~\texttt{integer\_of\_int32}), thus only allowing the injection of values fitting the destination type.
\end{itemize}
The \texttt{defensive} model is simple and efficient and is suitable for most cases except when precise modeling of machine arithmetic or bitwise operations is needed. For these purposes \textsc{Jessie} implements \texttt{modulo} integer model, which precisely models values of integral types as bitvectors.

Unfortunately, the integer model in \textsc{Jessie} can only be chosen once for the entire program analyzed using the corresponding pragma. In practice, however, it is desirable to be able to choose the appropriate integer model on a very fine-grained basis, down to every arithmetic operation. Consider the following example:
\begin{lstlisting}[caption={from Linux 4.12, \texttt{lib/string.c}},captionpos=b]
int strncasecmp(const char *s1, const char *s2, size_t len) {
 unsigned char c1, c2;
 if (!len) return 0;
 do {
   c1 = *s1++;
   c2 = *s2++;
   if (!c1 || !c2) break;
   if (c1 == c2) continue;
   c1 = tolower(c1);
   c2 = tolower(c2);
   if (c1 != c2) break;
 } while (--len);
 return (int)c1 - (int)c2;
}
\end{lstlisting}
Here in lines \texttt{5} and \texttt{6} the \texttt{modulo} integer model would be suitable as the cast from \texttt{\textbf{char}} to \texttt{\textbf{unsigned char}} may overflow and this is in line with the intention of the programmer. However we would also like a potential overflow to be detected if we accidentally change the return type of the function to \texttt{\textbf{char}}. So the \texttt{defensive} integer model is suitable to model the subtraction in line \texttt{13}.

To support such fine-grained integer model selection, we implemented an extension to the ACSL specification language with \emph{modulo arithmetic annotations}. The following new modulo arithmetic annotations were introduced:
\begin{itemize}
  \item for arithmetic operations: \texttt{+\textit{/*@\%*/}}, \texttt{--\textit{/*@\%*/}}, \texttt{*\textit{/*@\%*/}}, ...
  \item for compound assignments: \texttt{+=\textit{/*@\%*/}}, \texttt{--=\textit{/*@\%*/}}, \texttt{/=\textit{/*@\%*/}}, ...
  \item for prefix and postfix operators: \texttt{++\textit{/*@\%*/}}, \texttt{-- --\textit{/*@\%*/}}
  \item for explicit casts: \texttt{(\textbf{unsigned char})\textit{/*@\%*/}}, ...
  \item for modulo arithmetic in logic: \texttt{+\%}, \texttt{--\%}, \texttt{*\%}, ...
\end{itemize}

The integer model used to model both \texttt{defensive} (the default) and \texttt{modulo} arithmetic operations is a combined one. In this model, bounded integers are modeled as bitvectors with two injection/projection functions to/from the mathematical unbounded integers.

\sloppypar
Let's illustrate the encoding employed by the model on a sample arithmetic operation \texttt{+}, a bitwise operation \texttt{\&}, a relation \texttt{<}, and a sample bounded integer type \texttt{bint} (with injection function \texttt{to\_int}). The operation \texttt{+} \emph{in logic} is encoded simply as integer addition and has type $\texttt{int} \times \texttt{int} \to \texttt{int}$. The operation \texttt{+\%} \emph{in logic} is encoded as bitvector addition and has type \mbox{$\texttt{bint} \times \texttt{bint} \to \texttt{bint}$}. It is also augmented with axioms relating the operation to \texttt{+}, \eg $\forall a, b : \texttt{bint}.~\texttt{in\_bounds}\big(\texttt{to\_int}(a) \mathop{\texttt{+}} \texttt{to\_int}(b)\big) \implies \texttt{to\_int}(a \mathop{\texttt{+\%}} b) = \texttt{to\_int}(a) \mathop{\texttt{+}} \texttt{to\_int}(b)$. The operation \texttt{\&} \emph{in logic} is encoded as bitwise conjunction with the same type as \texttt{+\%}. The relation \texttt{<} \emph{in logic} is encoded as either bitwise or integer relation depending on the type of arguments. The bitwise relations is augmented with an axiom relating it to the integer one. The operation $a \mathop{\texttt{+}} b$ \emph{in code} is encoded as an abstract operation (\texttt{\textbf{val}} in \textsc{Why3ML}) on bitvectors with precondition requiring $\texttt{in\_bounds}(\texttt{to\_int}(a) \mathop{\texttt{+}} \texttt{to\_int}(b))$ end ensuring two postconditions: $\texttt{\textit{result}}=a \mathop{\texttt{+\%}} b$ and $\texttt{to\_int}(\texttt{\textit{result}})=\texttt{to\_int}(a) \mathop{\texttt{+}} \texttt{to\_int}(b)$. The operation $a \mathop{\texttt{+\textit{/*@\%*/}}} b$ \emph{in code} is also encoded as an abstract operation with no precondition and two postconditions: $\texttt{\textit{result}}=a \mathop{\texttt{+\%}} b$ and $\texttt{to\_int}(\texttt{\textit{result}})=\texttt{norm}(\texttt{to\_int}(a) \mathop{\texttt{+}} \texttt{to\_int}(b))$, where \texttt{norm} stands for an expression for range normalization involving axiomatization of modulo arithmetic. The operation \texttt{\&} \emph{in code} is an abstract operation with a straightforward postcondition $\texttt{\textit{result}}=a \mathop{\texttt{\&}} b$. Finally, the predicate \texttt{<} \emph{in code} is an abstract operation with two postconditions $\texttt{\textit{result}} \iff a \mathop{\texttt{<}} b$ and $\texttt{\textit{result}} \iff \texttt{to\_int}(a) \mathop{<} \texttt{to\_int}(b)$. Other operations are represented similarly.

This encoding enables construction of more expressive and predictable models while avoiding direct use of any interpretation for function \texttt{to\_int}, which usually can't be efficiently handled by the solvers. On the other hand, the use of quantified axioms significantly reduces both predictability and performance of the solvers. This can be potentially addressed by either adding some preliminary instantiation step or implementing similar support for the necessary operations as an SMT theory directly in the solver (by converting axioms into inference rules of the theory).

Lastly, let's demonstrate some practical capabilities of this integer model, even in the naive implementation, with an example proof of a bit-twiddling trick for computing average of two unsigned integers:
\begin{lstlisting}[captionpos=b]
//@ ensures \result ≡ (a + b) / 2;
unsigned average(unsigned a, unsigned b)
{
  /*@ ghost unsigned long long result1 =
      (a ^ b) + ((unsigned long long) (a & b) << 1ULL); */
  /*@ ghost unsigned long long result2 =
      (unsigned long long) a + b; */
  //@ assert result1 ≡ result2;
  return (a & b) + ((a ^ b) >> 1U);
}
\end{lstlisting}
Here the expressions in the ghost code trigger succinct instantiation of necessary lemmas relating bitwise and integer interpretations of bounded integers (through the double post-conditions of the corresponding \textsc{Why3} operations).

The use of such a combined model and the introduction of new fine-grained modulo arithmetic annotations allowed us to significantly simplify the specification and verification of many functions included in this study.

\section{Formal Specifications}\label{sec:specs}

We were guided by several techniques in the development of specifications: the use of excessive specifications (explicit specifications and specifications that establish the correspondence with a logical function), the development of specifications based on source code, and the context of function calls.

The results described in~\cite{openbsdverification} show that the development of a function contract, based exclusively on documentation is difficult: almost always, at the proof stage we have to rewrite the specification based on the source code. This approach is also explained by the fact that in this work we develop specifications on the complete code. Linux code is not written in accordance with a certain set of formal specifications. Also, the kernel does not have documentation for a lot of functions. We intentionally did not follow the standard documentation (man pages) for such functions, since their implementation in the kernel can differ from the others (for example, from implementation in the standard library), and the documentation is incomplete and may contain inaccuracies~\cite{openbsdverification}.


\begin{lstlisting}[caption={\texttt{strnlen} contract},label={lst:strnlenverker},captionpos=b]
/*@ predicate valid_strn(char *s, size_t cnt) =
     (∃ size_t n; n<cnt ∧ s[n] ≡ '\0' ∧ \valid(s+(0..n))) ∨
     \valid(s+(0..cnt));
    requires valid_strn(s, cnt);
    assigns \nothing;
    ensures \result ≡ strnlen(s, cnt);
    behavior null_byte:
      assumes ∃ ℤ i; 0 ≤ i ≤ cnt ∧ s[i] ≡ '\0';
      ensures s[\result] ≡ '\0';
      ensures ∀ ℤ i; 0 ≤ i < \result ⇒ s[i] ≢ '\0';
    behavior cnt_len:
      assumes ∀ ℤ i; 0 ≤ i ≤ cnt ⇒ s[i] ≢ '\0';
      ensures \result ≡ cnt;
    complete behaviors; disjoint behaviors;*/
size_t strnlen(const char *s, size_t cnt);
\end{lstlisting}

Following this approach, the specifications for some functions have a slightly more detailed view. For example, for \texttt{strn}* functions (see~\Cref{lst:strnlenverker,lst:strncmpverker}) we do not require the presence of the string's end marker. In the \texttt{strnlen}'s precondition (see~\Cref{lst:strnlenverker}), it is assumed that the string should be valid until the minimum of the string's length (if there is one) and the second argument of the function \texttt{strnlen}. The return value is explicitly specified in the postcondition. In the \texttt{strncmp} case (see~\Cref{lst:strncmpverker}), there are also no restrictions on the fact that the input strings must contain a zero byte. This leads to the point where it is necessary to explicitly describe the behavior of the function when the input strings with end markers differ in length. We tried to maximally weaken the preconditions and strengthen the postcondition in order to test the instruments of deductive verification, the expressiveness of the ACSL language, and the capabilities of solvers.



\begin{lstlisting}[caption={\texttt{strncmp} contract},label={lst:strncmpverker},captionpos=b,breaklines=false,breakautoindent=false,linewidth=200em]
/*@ requires valid_strn(cs, cnt) ∧ valid_strn(ct, cnt);
    assigns \nothing;
    ensures \result ≡ -1 ∨ \result ≡ 0 ∨ \result ≡ 1;
    behavior equal:
      assumes cnt ≡ 0 ∨ (cnt > 0  ∧
        (∀ ℤ i; 0 ≤ i<strnlen(cs,cnt) ⇒ (cs[i] ≡ ct[i])) ∧
        strnlen(cs, cnt) ≡ strnlen(ct, cnt));
      ensures \result ≡ 0;
    behavior len_diff:
      assumes cnt > 0;
      assumes ∀ ℤ i; 0 ≤ i<min(strnlen(cs,cnt),strnlen(ct,cnt))
                    (*@@*) ⇒ cs[i] ≡ ct[i];
      assumes strnlen(cs, cnt) ≢ strnlen(ct, cnt);
      ensures strnlen(cs,cnt)<strnlen(ct,cnt) ⇒ \result ≡ -1;
      ensures strnlen(cs,cnt)>strnlen(ct,cnt) ⇒ \result ≡ 1;
    behavior not_equal:
      assumes cnt > 0;
      assumes ∃ ℤ i; 0 ≤ i<strnlen(cs, cnt) ∧ cs[i] != ct[i];
      ensures ∃ ℤ i; 0 ≤ i<strnlen(cs, cnt) ∧
        (∀ ℤ j; 0 ≤ j<i ⇒ cs[j] ≡ ct[j]) ∧
        cs[i] ≢ ct[i] ∧
        ((u8 %)cs[i]<(u8 %)ct[i]? \result ≡ -1: \result ≡ 1);
    complete behaviors; disjoint behaviors;*/
int strncmp(const char *cs, const char *ct, size_t cnt);
\end{lstlisting}

\subsection{Logic Functions}

The specifications are redundant for some functions. In fact, they describe a function's behavior in two different ways. For example, \texttt{strlen} specification consists of the usual functional requirements and the requirement for the correspondence between the returned value and the logical function. This approach is motivated by the fact that the logic function \texttt{strlen} is convenient to use in specifications of other functions, \eg \texttt{strcmp} (and a logical function that describes the behavior of the function \texttt{strcmp}~--- when describing the functional requirements for \mbox{\texttt{strcpy}}). All the basic properties of logic functions are specified by means of axioms and lemmas. The lemmas were not proved at the first stage presented in this paper only contradiction checks were performed\footnote{Since then we proved all the lemmas using techniques of auto-active verifivation~\cite{autoactive,adaredblack}, in particular, \emph{lemma functions}~\cite{quickverifast}. This work is available at~\cite{url:verkerlemmas}.}. However, such specifications do not suit all situations. For example, in the general case, they cannot be translated by E-ACSL~\cite{eacsl} as executable specifications. Therefore, for functions with an associated logical function, the ``usual'' specifications were also developed.

A logical function can be associated with a C function (one-to-one) only if the last one is ``pure''. A logical function is useful for developing specifications of other C-functions. For example, in postconditions of \texttt{memcpy}, you can express the equality of \texttt{src} and \texttt{dest} by calling the \texttt{memcmp} logical function.

\section{Open Issues}\label{sec:issues}

At the specification level, the authors faced many problems related to significant inaccuracies in the modeling of pointer operations, as well as the insufficient level of ACSL language support by the tools.

Thus, for the \texttt{memmove} function, there is the VC, which states that the \texttt{dest} and \texttt{src} pointers should lie in the same allocated memory block. This is necessary in order for the result of their comparison to be determined by the standard~\cite{c11}. Recall how the \texttt{memmove} function works: it copies a memory area of \texttt{n} bytes from the \texttt{src} address to \texttt{dst}, provided that the two memory regions can either overlap or be disjoint. To implement the latter condition, the function performs an ordinal comparison of the \texttt{dest} and \texttt{src}. In that case, if \texttt{dest} is located before \texttt{src} the byte-by-byte copy from the beginning of \texttt{src} is performed (thus, if the regions overlap, already copied part will be overwritten); if \texttt{dest} is located after \texttt{src}, then copying is performed starting at the end of the \texttt{src} memory region.

The memory model implemented in \textsc{AstraVer} plugin allows arithmetic operations on pointers (in \texttt{memmove} this is a comparison implemented through the difference between pointers) only when the pointers belong to the same allocated memory block. For \texttt{memmove}, this is not necessarily the case. If we state in the specification contract that \texttt{src} and \texttt{dest} may belong to different allocated memory blocks, then it is impossible to prove the VC that states that they should belong to the same memory block. The unproved VC is reflected in the results~(\Cref{table:bench}). Although comparison of pointers to different memory blocks is the undefined behavior in ACSI C, the comparison can be made defined by casting the pointers to the corresponding underlying integral type. Yet adding such casts is in odds with the goal of the presented work (verifying the functions without modifications) and also not currently supported by the \textsc{AstraVer} plugin.

The \texttt{strcat} function concatenates two strings by appending the \texttt{src} to the \texttt{dest}. To do this, the end of the \texttt{dest} string is determined first. Then the \texttt{src} string is copied in the same way as in \texttt{strcpy}. In order to prove the VCs that state the safety of memory operations in this function, it was enough to require the validity of the strings \texttt{src} and \texttt{dest} and sufficient memory behind the end of \texttt{dest} to accommodate the contents of \texttt{src}. However, proving the functional correctness of the implementation, it was revealed that it is necessary to formulate an additional requirement stating that the sum of the string's lengths fits the \texttt{size_t} type. The function is implemented through the pointers iteration. Therefore, the ability to prove the memory operation safety without the last requirement in the function means that the \texttt{AstraVer} memory model does not take into account the possibility of pointer overflow.


It was required to change the code of two functions to prove their correctness. Despite the fact that we want to minimize code changes, in two cases we cannot fully prove correctness without code modification. The functions \texttt{memset} and \texttt{strcmp} use the implicit type cast with overflow. \texttt{memset} casts \texttt{int} to \texttt{char}, and \texttt{strcmp} casts implicitly \texttt{char} to \texttt{unsigned char}. To mark these overflows as intentional it was required to make the casts explicit. Our ACSL extension with modulo arithmetic annotations still lacks the corresponding construction (\eg \mbox{\texttt{\textit{/*@(\textbf{unsigned int} \%)*/}}}) for implicit casts.



At the specification level, tools do not support the use of predicates in definitions of logical functions or predicates as first arguments of the ternary operator in lemmas and axioms. Because of this, it is sometimes difficult to give an explicit definition of a logical function, and we have to use an axiomatic (implicit) definition. This drawback prevents the explicit definition of the logical functions for \texttt{skip_spaces}, \texttt{strcspn}, \texttt{strpbrk}, and \texttt{strspn}.

Functions from the file \texttt{ctype.h} (\texttt{isspace}, \texttt{isdigit}, \texttt{isalnum}, \texttt{isgraph}, \texttt{islower},~...) are defined as macros that operate on the array \texttt{\_ctype} of 256 bytes, which specifies the belonging of each character to a particular class. To simplify the verification task, these macros have been replaced by inline functions: verification tools do not allow the writing of specifications for macros, only for functions. The \texttt{\_ctype} array was redefined as a string (string initialization is translated into model axioms) because the global array initialization is not translated into the WhyML model. However, it was not possible to prove the correspondence of functions from the \texttt{ctype.h} file to their specifications even after the described transformations: solvers cannot cope with the proof when the model has an axiomatic definition of the \texttt{\_ctype} array 256 characters long.

\section{Evaluation of Solvers}\label{sec:evaluation}

\textsc{AstraVer} translates \textsc{Frama-C}'s internal representation into the program model in WhyML~\cite{why3ide}, based on the memory model and semantics of operations with integers. The \textsc{Why3} tool generates VCs for a WhyML program and converts them into an input for solvers. \textsc{Why3} supports a number of solvers, such as \textsc{Alt-Ergo}, \textsc{CVC3}, \textsc{CVC4}, \textsc{Z3}, \textsc{Spass}, \textsc{EProver}, \textsc{Simplify} and others. \textsc{Why3} also supports transformations of VCs, for example, splitting conjunctions into separate conditions.

\textsc{Alt-Ergo} (1.30) and \textsc{CVC4} (1.4) SMT solvers are able to discharge all VCs generated (except for the one for \texttt{memmove}). However, it is interesting to evaluate other solvers on the given benchmark. For that purpose we conduct an experiment using the following system configuration: CPU~---~\mbox{AMD FX-8120} (Eight-Core Processor), RAM~---~16GB, time limit~---~60 seconds, memory limit~---~6000MB, OS~---~GNU/Linux (kernel: 4.12.12 (smp preempt) \mbox{x86_64}), software (from \textsc{AstraVer} repository): \textsc{Why3} (0.87.3+git), \mbox{\textsc{Frama-C}} (\mbox{Silicon-20161101}), \textsc{Jessie2} (alpha3).

\subsection{VC transformation strategy}
To put all solvers in similar conditions all VCs were transformed by \textsc{Why3} using the following strategy:
\begin{enumerate}
\item Split goal by conjuncts (\texttt{split\_goal\_wp}) repeatedly until fixed point.
\item Inline definition of all logical symbols (\texttt{inline\_all}).
\item Split goal by conjuncts (\texttt{split\_goal\_wp}) repeatedly until fixed point.
\item Skolemize goal (\texttt{introduce\_premises}).
\end{enumerate}

If there are many predicates with long dependency chains, the \texttt{inline\_all} transformation makes the work of the solvers more difficult. This is not the case for the given benchmark and experiments have shown the positive impact of this transformation. The addition of \texttt{introduce\_premises} transformation also comes from preliminary experiments demonstrating that solvers work better with formulas of the form $f(x) \land \neg g(x)$ than with ones of the form $\neg \forall x \ldotp~ f(x) \implies  g(x)$. Otherwise, the strategy tries to split the VC into the smallest possible conjuncts.

During the development of specifications, the strategy is not applied by default. Only some of the transformations are applied if solvers fail to discharge VCs by themselves.

\begin{table}[th!]
\resizebox{\textwidth}{!}{%
\begin{tabular}{|l||l|lr|lr|lr|lr|lr|lr|lr|}
\hline
\multicolumn{1}{|c||}{\textbf{Function}} & \multicolumn{1}{c|}{\textbf{VC}} & \multicolumn{2}{c|}{\textbf{Alt-Ergo}} & \multicolumn{2}{c|}{\textbf{CVC3}}  & \multicolumn{2}{c|}{\textbf{CVC4}} & \multicolumn{2}{c|}{\textbf{CVC4}} & \multicolumn{2}{c|}{\textbf{Eprover}}   & \multicolumn{2}{c|}{\textbf{Spass}} & \multicolumn{2}{c|}{\textbf{Z3}}    \\
\multicolumn{1}{|c||}{\textbf{}}         & \multicolumn{1}{c|}{\textbf{}}   & \multicolumn{2}{c|}{\textbf{1.30}}     & \multicolumn{2}{c|}{\textbf{2.4.1}} & \multicolumn{2}{c|}{\textbf{1.4}}  & \multicolumn{2}{c|}{\textbf{1.5}}  & \multicolumn{2}{c|}{\textbf{1.9.1-001}} & \multicolumn{2}{c|}{\textbf{3.9}}   & \multicolumn{2}{c|}{\textbf{4.5.0}} \\
\hline
\hline
                & total & vc         & atime                  & vc                   & atime                  & vc                    & atime                  & vc                     & atime                 & vc   & atime                  & vc                  & atime                   & vc   & atime                  \\
\hline
\_parse\_integ. & 282   & 276        & 0.10                   & 280                  & 0.83                   & \checkmark            & 0.18                   & \checkmark             & 0.10                  & 212  & 0.24                   & \cellcolor{red} 197 & \cellcolor{brown} 1.69  & 279  & \cellcolor{cyan} 0.06  \\
check\_bytes8   & 50    & 49         & 0.55                   & 49                   & \cellcolor{cyan} 0.09  & 49                    & \cellcolor{cyan} 0.09  & \checkmark             & 0.11                  & 38   & 1.76                   & \cellcolor{red} 31  & \cellcolor{brown} 8.38  & 36   & 1.52                   \\
kstrtobool      & 1096  & \checkmark & \cellcolor{cyan} 0.05  & \checkmark           & 0.08                   & \checkmark            & 0.10                   & \checkmark             & 0.09                  & 1006 & 0.13                   & \cellcolor{red} 937 & \cellcolor{brown} 0.38  & 1065 & 0.15                   \\ \hline
memchr          & 39    & \checkmark & \cellcolor{brown} 6.05 & \cellcolor{red} 11   & 0.22                   & \checkmark            & 0.37                   & \checkmark             & 0.15                  & 31   & 2.58                   & \cellcolor{red} 11  & 5.73                    & 29   & \cellcolor{cyan} 0.12  \\
memcmp          & 60    & 58         & 0.13                   & \checkmark           & 0.15                   & 58                    & \cellcolor{cyan} 0.10  & \checkmark             & \cellcolor{cyan} 0.10 & 49   & 0.51                   & \cellcolor{red} 36  & \cellcolor{brown} 4.45  & 55   & 0.15                   \\
memcpy          & 43    & \checkmark & 4.18                   & \checkmark           & 0.35                   & \checkmark            & 0.16                   & \checkmark             & 0.14                  & 30   & 1.05                   & \cellcolor{red} 16  & \cellcolor{brown} 6.85  & 30   & \cellcolor{cyan} 0.06  \\
memmove         & \textbf{93}*(92)    & 90         & 3.94                   & \checkmark           & 0.88                   & 87                    & \cellcolor{cyan} 0.16  & \checkmark             & 0.18                  & 63   & 0.95                   & \cellcolor{red} 43  & \cellcolor{brown} 11.87 & 68   & 0.30                   \\
memscan         & 47    & 46         & 0.07                   & \checkmark           & 0.10                   & \checkmark            & 0.09                   & \checkmark             & 0.09                  & 41   & 0.59                   & \cellcolor{red} 34  & \cellcolor{brown} 4.55  & 42   & \cellcolor{cyan} 0.06  \\
memset          & 27    & 26         & 5.02                   & 14                   & 0.19                   & \checkmark            & 0.19                   & \checkmark             & 0.16                  & 19   & 3.82                   & \cellcolor{red} 12  & \cellcolor{brown} 11.12 & 18   & \cellcolor{cyan} 0.08  \\ \hline
skip\_spaces    & 34    & 30         & 0.76                   & 32                   & \cellcolor{brown} 1.96 & \checkmark            & 0.51                   & 33                     & 0.14                  & 27   & 0.70                   & \cellcolor{red} 24  & 0.34                    & 30   & \cellcolor{cyan} 0.09  \\
strcasecmp      & 58    & 50         & 0.43                   & 52                   & 1.65                   & 57                    & 0.79                   & \checkmark             & 0.53                  & 43   & \cellcolor{cyan} 0.28  & \cellcolor{red} 35  & \cellcolor{brown} 2.85  & 49   & 0.49                   \\
strcat          & 73    & 68         & 0.58                   & 66                   & 2.16                   & \checkmark            & 1.13                   & 71                     & \cellcolor{cyan} 0.17 & 54   & \cellcolor{brown} 2.56 & \cellcolor{red} 39  & 0.67                    & 60   & 0.94                   \\
strchr          & 43    & 35         & \cellcolor{brown} 4.57 & \cellcolor{red} 23   & 0.17                   & \checkmark            & 0.23                   & \checkmark             & 0.22                  & 31   & 1.03                   & 24                  & 3.65                    & 32   & \cellcolor{cyan} 0.11  \\
strchrnul       & 46    & 42         & 2.07                   & 37                   & 0.26                   & \checkmark            & 0.19                   & \checkmark             & \cellcolor{cyan} 0.16 & 40   & 1.91                   & \cellcolor{red} 31  & \cellcolor{brown} 2.27  & 39   & 0.31                   \\
strcmp          & 60    & 51         & \cellcolor{brown} 1.76 & \cellcolor{red} 16   & 0.60                   & \checkmark            & 1.75                   & 59                     & 1.08                  & 47   & 1.05                   & 36                  & 1.65                    & 47   & \cellcolor{cyan} 0.10  \\
strcpy          & 46    & 43         & 1.33                   & 45                   & 0.66                   & \checkmark            & 0.48                   & \checkmark             & \cellcolor{cyan} 0.17 & 33   & 1.13                   & \cellcolor{red} 26  & 0.65                    & 39   & \cellcolor{brown} 1.43 \\
strcspn         & 78    & 68         & 0.38                   & 69                   & 0.37                   & 74                    & \cellcolor{brown} 2.95 & \cellcolor{green} 75   & 1.82                  & 58   & 1.85                   & \cellcolor{red} 46  & 1.68                    & 61   & \cellcolor{cyan} 0.11  \\
strlcpy         & 84    & 82         & 0.15                   & 82                   & \cellcolor{cyan} 0.14  & \checkmark            & 1.08                   & \checkmark             & 0.24                  & 67   & 1.20                   & \cellcolor{red} 52  & \cellcolor{brown} 1.74  & 78   & 0.42                   \\
strlen          & 26    & \checkmark & 1.12                   & 24                   & 0.12                   & \checkmark            & 0.16                   & \checkmark             & 0.23                  & 19   & \cellcolor{brown} 3.36 & \cellcolor{red} 14  & 2.96                    & 21   & \cellcolor{cyan} 0.08  \\
strnchr         & 49    & 38         & \cellcolor{brown} 4.44 & \cellcolor{red} 19   & 0.23                   & 46                    & 3.34                   & \checkmark             & 0.72                  & 35   & 2.57                   & 24                  & 1.56                    & 27   & \cellcolor{cyan} 0.09  \\
strncmp         & 102   & 81         & \cellcolor{brown} 2.57 & \cellcolor{red} 25   & \cellcolor{cyan} 0.25  & 94                    & 2.39                   & \cellcolor{green} 99   & 2.32                  & 76   & 1.06                   & 55                  & 2.56                    & 76   & 0.57                   \\
strnlen         & 44    & 39         & 1.91                   & 42                   & 1.04                   & 39                    & 1.23                   & \checkmark             & 1.31                  & 31   & 2.40                   & \cellcolor{red} 26  & \cellcolor{brown} 5.52  & 32   & \cellcolor{cyan} 0.08  \\
strpbrk         & 70    & 57         & 0.64                   & 58                   & 1.54                   & 62                    & \cellcolor{brown} 3.18 & \cellcolor{green} 67   & 1.57                  & 48   & 1.89                   & \cellcolor{red} 39  & 0.75                    & 53   & \cellcolor{cyan} 0.09  \\
strrchr         & 62    & 53         & 4.57                   & \cellcolor{red} 12   & 0.17                   & \checkmark            & 1.09                   & 60                     & 0.85                  & 46   & 2.33                   & 31                  & \cellcolor{brown} 4.67  & 46   & \cellcolor{cyan} 0.11  \\
strsep          & 62    & 60         & 0.25                   & 60                   & 0.09                   & \checkmark            & 0.19                   & \checkmark             & 0.15                  & 55   & 0.12                   & \cellcolor{red} 51  & \cellcolor{brown} 1.48  & 58   & \cellcolor{cyan} 0.06  \\
strspn          & 107   & 99         & 0.84                   & 100                  & 0.69                   & \cellcolor{green} 104 & 1.32                   & 103                    & 0.61                  & 89   & 1.37                   & \cellcolor{red} 75  & \cellcolor{brown} 1.59  & 91   & \cellcolor{cyan} 0.13  \\
\hline
TOTAL           & 2781  & 2645       & 0.90                   & 2454                 & 0.42                   & 2740                  & 0.61                   & \cellcolor{green} 2761 & 0.37                  & 2288 & 0.76                   & \cellcolor{red} 1945& \cellcolor{brown} 1.72  & 2461 & \cellcolor{cyan} 0.22  \\
\hline
\end{tabular}}
\caption[Solvers]{\textbf{Solvers.} Proofs Statistics (times are given in seconds)}
\label{table:bench}
\end{table}

\vspace{1em}

\begin{table}[th!]
	\resizebox{\textwidth}{!}{%
	\begin{tabular}{|lr|lr|lr|lr|lr|lr|lr|}
	\hline
	\multicolumn{2}{|c|}{\textbf{Alt-Ergo}} & \multicolumn{2}{c|}{\textbf{CVC3}}  & \multicolumn{2}{c|}{\textbf{CVC4}} & \multicolumn{2}{c|}{\textbf{CVC4}} & \multicolumn{2}{c|}{\textbf{Eprover}}   & \multicolumn{2}{c|}{\textbf{Spass}} & \multicolumn{2}{c|}{\textbf{Z3}}    \\
	\multicolumn{2}{|c|}{\textbf{1.30}}     & \multicolumn{2}{c|}{\textbf{2.4.1}} & \multicolumn{2}{c|}{\textbf{1.4}}  & \multicolumn{2}{c|}{\textbf{1.5}}  & \multicolumn{2}{c|}{\textbf{1.9.1-001}} & \multicolumn{2}{c|}{\textbf{3.9}}   & \multicolumn{2}{c|}{\textbf{4.5.0}} \\
	\hline
	mtime & uniq & mtime & uniq & mtime & uniq & mtime & uniq & mtime & uniq & mtime & uniq & mtime & uniq \\ \hline
	\hline
	58.75 & 1 &
	56.68 & 0 &
	57.97 & 7 &
	52.27 & 20 &
	47.80 & 0 &
	59.74 & 0 &
	26.74 & 0 \\
	\hline
\end{tabular}}
\caption[Solvers]{\textbf{Solvers.} Max Time and Number of Uniq Proofs}
\label{table:maxtime}
\end{table}


\subsection{Statistics}

\Cref{table:bench} presents the results of the evaluation. The first column contains the target function name the second one includes the number of VCs generated (safety and behavioral) after application of the transformation strategy. The rest of the table presents solver statistics: the amount of discharged VCs and the average time for successful runs.

\fboxsep0.8pt

The symbol \checkmark\  marks cases when a solver proved all VCs for the corresponding function. Maximal numbers of discharged VCs are highlighted in \colorbox{green}{green}. Minimal VC counts are highlighted by \colorbox{red}{red}. The minimal average time is highlighted in \colorbox{cyan}{cyan}, the maximal average time is highlighted in \colorbox{brown}{brown}.


\subsection{Discussion}
All VCs except one for \texttt{memmove} are successfully discharged by at least one of the solvers. The best result was achieved by \textsc{Alt-Ergo} and \textsc{CVC4}. This is expected as those solvers were most extensively used during the development and testing of the toolset.

\textsc{CVC4} 1.5 discharged the greatest number of VCs, while \textsc{Z3} required the smallest amount of time. This can be partially explained by the fact that we counted only successful proof attempts. \textsc{Z3} was able to prove fewer VCs than \textsc{Alt-Ergo} or \textsc{CVC4}.


\Cref{table:maxtime} presents maximal solving times for successful proof attempts and counts of unique proofs, \ie VCs that were only discharged by one of the solvers.


\section{Conclusion}\label{sec:conclusion}

This paper presents results from the development and evaluation of a deductive verification benchmark consisting of 26 unmodified Linux kernel library functions implementing conventional memory and string operations. Formal contracts of the functions were extracted from their source code and were represented in the form of preconditions and postconditions. The benchmark detected a number of problems with existing deductive verification toolchains. Some of the issues required only fixes in the tools, some of them led to the design of proposals to extend ACSL language, others were left open.

For example, two newly proposed ACSL constructs allowed us to successfully proof 11 more functions without modification of their source code. With these extensions, the authors have successfully and fully proved the correctness of 23 functions. Another 2 functions were proved after a minor modification of their source code, while the final one cannot be completely proved using existing memory model. Specifications of the benchmark contain $\approx$ 2.6 times as many lines as the source code of the library functions.

The source code of the benchmark and proof protocols are publicly available together with instructions describing how to reproduce the results~\cite{url:verkerresults}. The benchmark can be used for the testing and evaluation of deductive verification tools and the starting point for verifying other parts of Linux kernel. A possible next step is to extend the benchmark with other library functions (\eg bitwise operations).




\bibliographystyle{splncs03}
\bibliography{FRAMAC}

\end{document}